\documentclass[twocolumn]{svjour3}          
\smartqed  
\usepackage{graphicx}
\begin{document}

\title{Josephson coupling in junctions made of monolayer graphene grown on SiC}

\author{
B.~Jouault$^1$\and
S. Charpentier$^2$\and
D. Massarotti$^{3,4}$\and 
A. Michon$^5$\and
M. Paillet$^1$\and
J.-R. Huntzinger$^1$\and
A. Tiberj$^1$\and
A. Zahab$^1$\and
T. Bauch$^2$\and
P. Lucignano$^4$\and
A. Tagliacozzo$^{6,4,7}$\and
F. Lombardi$^2$\and
F. Tafuri$^{4,3}$}

\institute{
$^1$\at Universit\'e Montpellier-CNRS, Laboratoire Charles Coulomb UMR 5221, F-34095, Montpellier, France
\and
$^2$\at Chalmers University of Technology, SE-412 96 G\"oteborg, Sweden
\and
$^3$\at Dipartimento di Ingegneria Industriale e dell'Informazione, Seconda Universit\'a di Napoli, I-81031Aversa (CE), Italy
\and
$^4$\at
CNR-SPIN, Monte S. Angelo-Via Cintia, I-80126, Napoli, Italy
\and
$^5$\at
CRHEA - Centre de Recherche sur l'H\'et\'ero\'epitaxie et ses Applications,
CNRS, rue Bernard Gr\'egory, 06560 Valbonne, France
\and
$^6$\at
Dipartimento di Fisica, Universit\`a di Napoli "Federico II", Monte S. Angelo-Via Cintia, I-80126 Napoli, Italy
\and
$^7$\at
INFN, Laboratori Nazionali di Frascati, Via E.Fermi, Frascati, Italy
}

\date{\today}

\maketitle

\begin{abstract}
Chemical vapor deposition has proved to be successful
in producing graphene samples on silicon carbide (SiC) homogeneous at the centimeter scale in terms of Hall conductance quantization.
Here we report on the realization of co-planar diffusive Al/ mono\-layer gra\-phene/ Al junctions on the same graphene sheet, with separations between the electrodes down to 200 nm. Robust Josephson coupling, as the magnetic pattern of the critical current, has been measured for separations not larger than 300 nm. Transport properties are reproducible on different junctions and indicate that graphene on SiC substrates is a concrete candidate to provide scalability of hybrid Josephson graphene/\-super\-con\-ductor devices.
\end{abstract}

\maketitle

\section{Introduction}
The remarkable electrical properties of graphene, induced by the chiral nature of its charge carriers, rises  big expectations presently, of increasing functionalities in hybrid superconducting quantum  devices.
Josephson junctions in which the supercurrent flows through normal-conducting graphene across closely spaced Superconductors (S), have been realized up to now only  with exfoliated graphene. Usually the graphene flakes are  deposited on a  Si/SiO$_2$ substrate\cite{HeerscheSSC2007,Heersche2007Nature,DuPRB2008,OjedaPRB2009,Borzenets2011,Lee2011PRL,PopinciusPRB2012,Coskun2012PRL}.

Even if the achievements are for many respects spectacular at  present, the goal of realizing whatever simple hybrid circuit with a number of these junctions on the same graphene sheet is a long way off, because of the severe limits imposed by the present fabrication protocols. Every junction will have its own properties (and story), with no chance of controlling the electrodynamics and its functionality, when engineering the full circuit. 

In this work we propose to use Chemical Vapor Deposition (CVD) of graphene on large  silicon carbide (SiC) areas, which does not require subsequent exfoliation. We report on the characteristics of the fabricated  junctions, demonstrating that the high quality of the 
devices, patterned on a single graphene sheet, paves the way to finally taking on this limit, an utmost priority in view of any future application.

%

%

%
%
Graphene made by CVD on metals\cite{LiCVD} and graphene on SiC cover large surfaces and are homogeneous at the centimetre scale. Thus relatively simple lithography processes allow to obtain thousands of devices on the same wafer\cite{Kedzierski2008IEEE}.
The mobility of graphene grown by CVD on metals is mainly limited by defects introduced during the transfer process\cite{ChanNL2012}, whereas the mobility of graphene on SiC is essentially controlled by the carrier concentration. 
In graphene on SiC, the mobility can reach $\mu \approx  45,000$ cm$^2$V$^{-1}$s$^{-1}$ at $T$=2 K and at carrier concentrations $2 \times 10^{10}$ cm$^{-2}$\cite{Tanabe2011PRB}.
This corresponds to a mean free path $l \approx$ 80 nm.
%
%
Additionally, the quantum Hall effect of graphene on SiC is remarkably precise even at magnetic fields as low as $B=3.5$ T\cite{Lafont2015ncom,Ribeiro2015nnano}, which makes graphene on SiC the material of choice for the next generation of easy-to-use quantum electrical standards.
By combining Josephson arrays supporting relatively large magnetic fields and Hall bars made of graphene, complex metrological devices like quantum current standards could be designed on the same wafer\cite{Poirier2014JAP,KomatsuPRB2012}.
Besides, the interface and/or the surface of graphene on SiC can be modified. Various atomic species like oxygen, hydrogen, calcium, can be intercalated or deposited  whereas it is well established that intercalated graphite or few layer graphene become superconductive\cite{LiAPL2013,Ludbrook2015,Tiwari2015}.
Thus, CVD graphene on SiC is also a good platform to study graphene-based superconductivity\cite{HanNP2014}.

\section{Sample Characterization and preparation}

\begin{figure}
\begin{center}
\includegraphics[width=0.9 \linewidth]{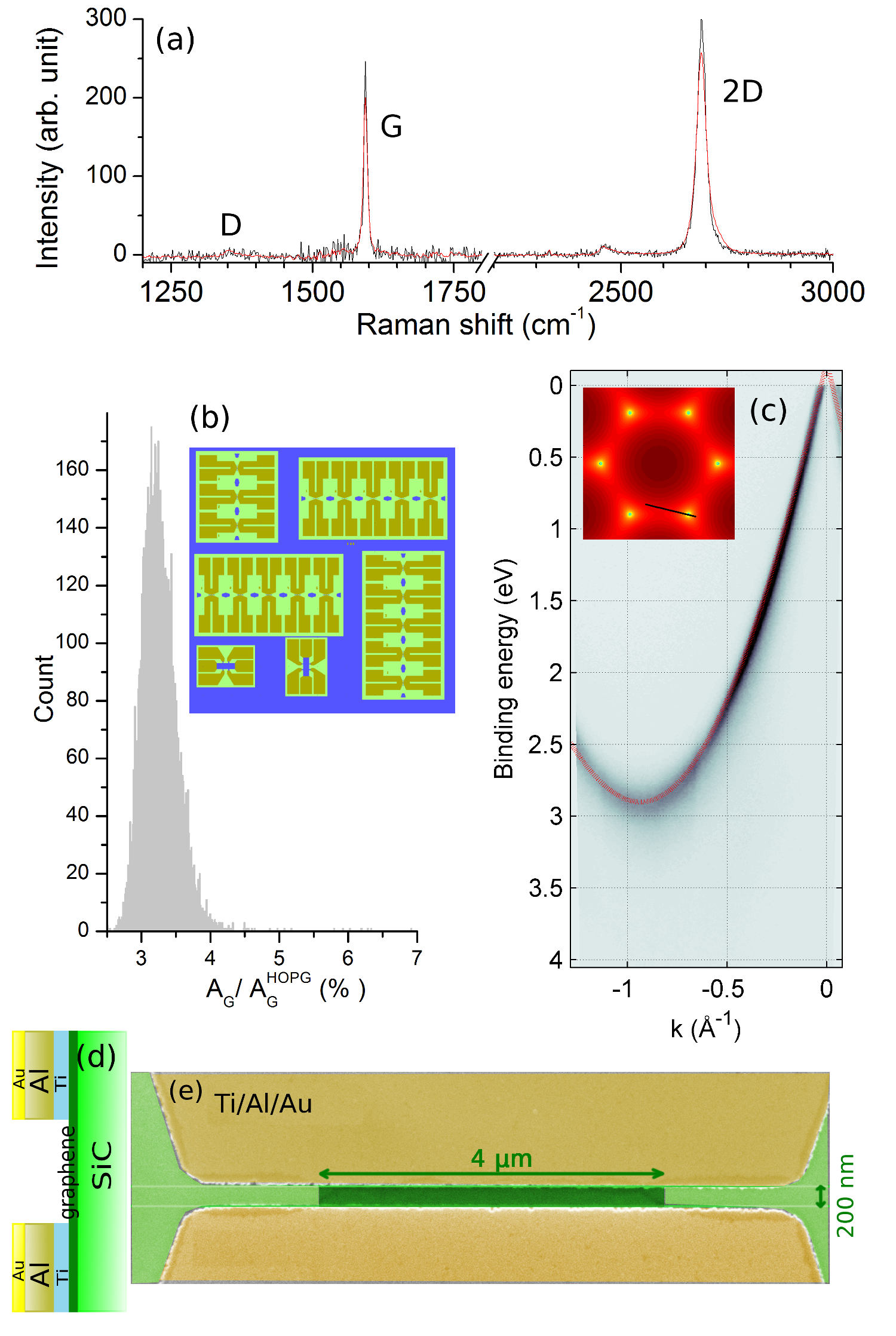}
\caption{(a) Raman spectra (with SiC background subtracted) obtained on a given location (black line) and averaged over a $24$ $\mu$m $\times 24$ $\mu$m area (red line). (b) Histogram of the integrated intensity $A_G$ of the G Raman peak, normalized over the integrated intensity $A_G^{HOPG}$ of the $G$ peak of a Highly Oriented Pyrolytic Graphite (HOPG) reference. 98\% of the surface gives a value close to 3.2 \%, which is close to expectations for
a monolayer graphene (see Ref.~\cite{Camara2009PRB}).  (c) Angle Resolved Photoemission Spectroscopy (ARPES) Intensity map of the sample surface. The red dashed line is the theoretical dispersion of the valence band of a monolayer graphene, according to Ref.~\cite{CastroNeto2009RMP}. The inset is a colormap of the graphene conduction band in the reciprocal space. The ARPES map is taken along the direction indicated by a black solid line. (d) Sketch and (e) scanning-electron micrograph of a $L\approx 200$ nm short junction with artificial colors. The sample structure after lithography is indicated in the inset of panel (b) for completeness (total size $5$ mm  $\times$ $5$ mm, green: SiC, blue: graphene, dark yellow: contacts).
}
\label{fig:analyse}
\end{center}
\end{figure}


Graphene has been grown by propane-\-hydrogen CVD\cite{michon10} on the Si face of a semi-insulating $0.24^o$-off-axis 6H-SiC substrate purchased from TankeBlue, using conditions similar to those used in Ref.~\cite{Jabakhanji2014PRBa}.
The growth is done in one minute, at a temperature of 1550~$^{\circ}$C under a 800 mbar atmosphere made of argon, propane and hydrogen (3 slm argon, 10 slm hydrogen, 5 sccm propane). The substrate is then cleaved to obtain several samples of total size $6$ mm $\times$ $6$ mm.
Fig.~\ref{fig:analyse} resumes the structural characteristics of the graphene samples. Raman spectra, obtained after the growth, are reported in panel (a). 
For 98\% of the surface, the integrated intensity of the $G$ peak, normalized with respect to a graphite reference, is about $3.2$\%, see panel (b). This is very close to the experimental value reported for a graphene monolayer\cite{Camara2009PRB}.
By contrast, bilayer graphene would yield about twice this intensity\cite{Camara2009PRB}. 
ARPES measurements, carried out on the Cassiop\'ee beamline at the synchrotron radiation facility SOLEIL, reveals that the sample is a $p$-doped monolayer graphene, see panel(c).
On a similar growth, magnetotransport measurements also revealed the half-integer quantum Hall effect specific to monolayer graphene\cite{Jabakhanji2014PRBa}.
The ARPES measurements are done under ultra-high vacuum, 
after an initial degassing performed at 500~$^o$C.
On the silicon-face of SiC, the graphene is usually $n$-doped, 
because of the presence of the so-called carbon-rich interface.
Here, the $p$-doping is the experimental signature that this carbon-rich interface has been neutralized and replaced by an hydrogenated interface between graphene and SiC\cite{Jabakhanji2014PRBa}. 
%

%
Van der Pauw and Hall measurements, done before lithography on the $6$ mm $\times$ $6$ mm sample, give a carrier concentration $p \approx$ $6.2 \times 10^{12}$ cm$^{-2}$ and a mobility $\mu \approx 1,100$  cm$^2\cdot$V$^{-1}\cdot$s$^{-1}$ at room temperature under ambient atmosphere.
This corresponds to a sheet resistivity $\rho= 920$ $\Omega$, with a form factor $f=0.9$ which evidences a small anisotropy of conductivity linked to the SiC steps. 
The low mobility is typical for graphene on SiC which is not compensated. In principle, mobility higher by one order of magnitude can be obtained by an appropriate gating which would reduce the carrier concentration\cite{Tanabe2011PRB}.
As both ARPES and Hall measurements give similar carrier concentration,  the atmospheric contamination plays a minor role and is not the main source of doping, which is likely to be due to the interface.

The Josephson devices are fabricated by conventional e-beam lithography. The graphene junctions are patterned by oxygen plasma. Then, the contacts are deposited, see Fig.~\ref{fig:analyse}(d,e). They consist of a 5 nm interfacial layer of titanium, a 80 nm thick aluminium layer and a 3 nm thick gold layer. The gaps $L$ between the electrodes ranges from 200 nm to 600 nm. The width is fixed at $W = 4$ $\mu$m.
%
A total of twelve junctions have been defined, with two orientations with respect to the substrate.
The samples have been thermally anchored to the $^3$He pot of a $^3$He cryostat (Heliox VL Oxford Instruments) and four probes electrical measurements have been performed. The cryostat is equipped with a room temperature electromagnetic interference filter stage followed by low pass RC filters anchored at 1.5 K and by copper powder filters stage anchored at the sample stage\cite{luigiPRB}. Standard measurements of current-voltage $V(I)$ characteristics as a function of temperature and magnetic field along with $R(T)$ curves with a bias current of about 5 nA have been performed. In addition, conductance spectra $dI/dV (V)$ have been measured by superimposing a low amplitude (a few nA) sinusoidal signal with frequency of about 30 Hz to a triangular slow ramp, with a frequency of about 1 mHz, and by reading the response from the sample by using the lock-in technique.
 
\section{Results}
\begin{figure}[h]
\begin{center}
\includegraphics[width=0.9 \linewidth]{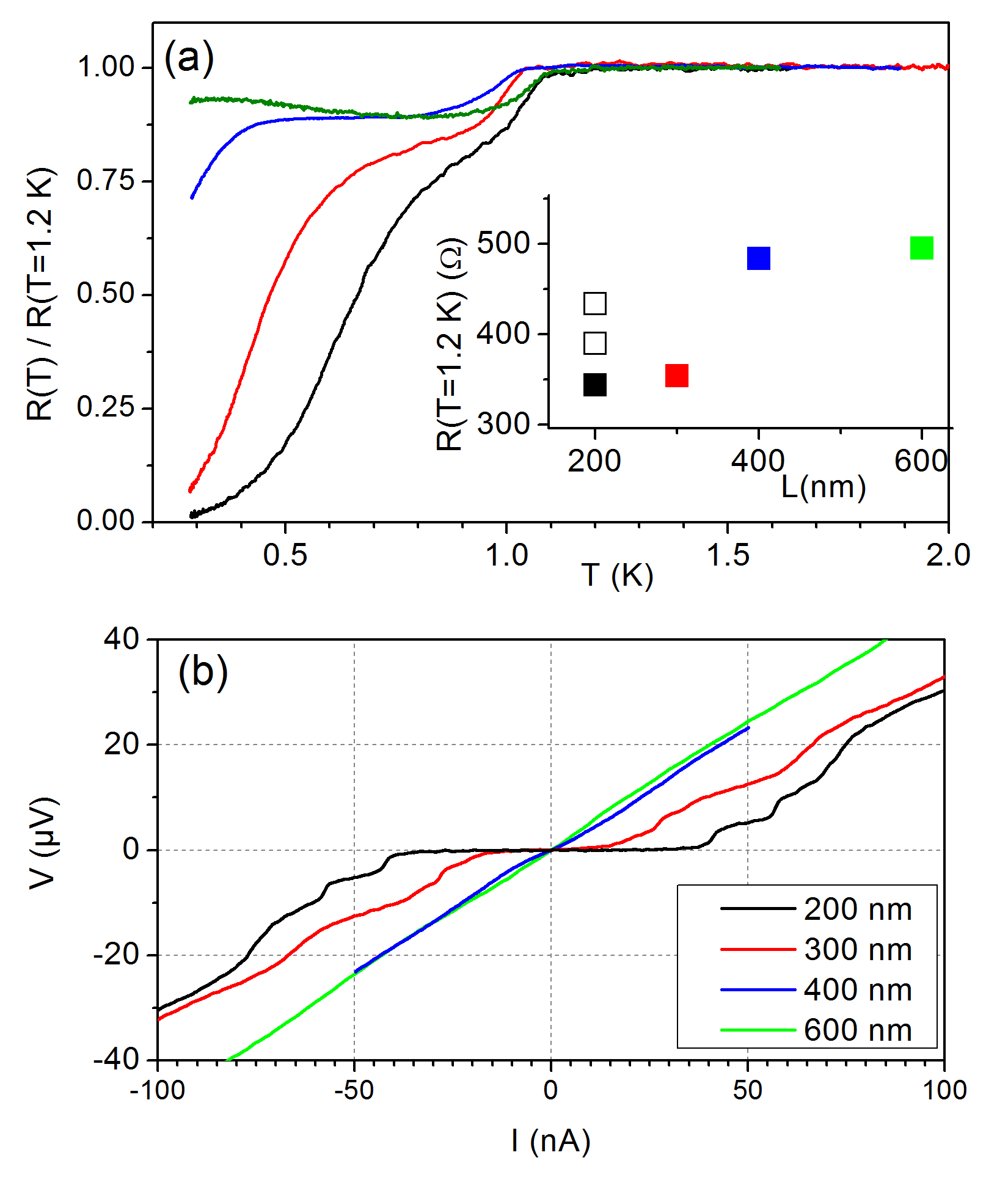}
\caption{(a) Temperature dependence of the resistance $R(T)$ for four junctions of lengths $L\approx 200, 300, 400$ and $600$ nm, normalized by their resistance at $T= 1.2$ K. The inset represents the resistance at $T= 1.2$ K of the same four junctions, plotted as a function of their width as filled squares with the same colour code. Two additional junctions of width $L\approx 200$ nm are also plotted as open squares. (b) V(I) characteristics of the four junctions presented in panel (a), at the base temperature of $T= 280$ mK.}
\label{fig:G0}
\end{center}
\end{figure}

\begin{figure}[h]
\begin{center}
\includegraphics[width=1.0 \linewidth]{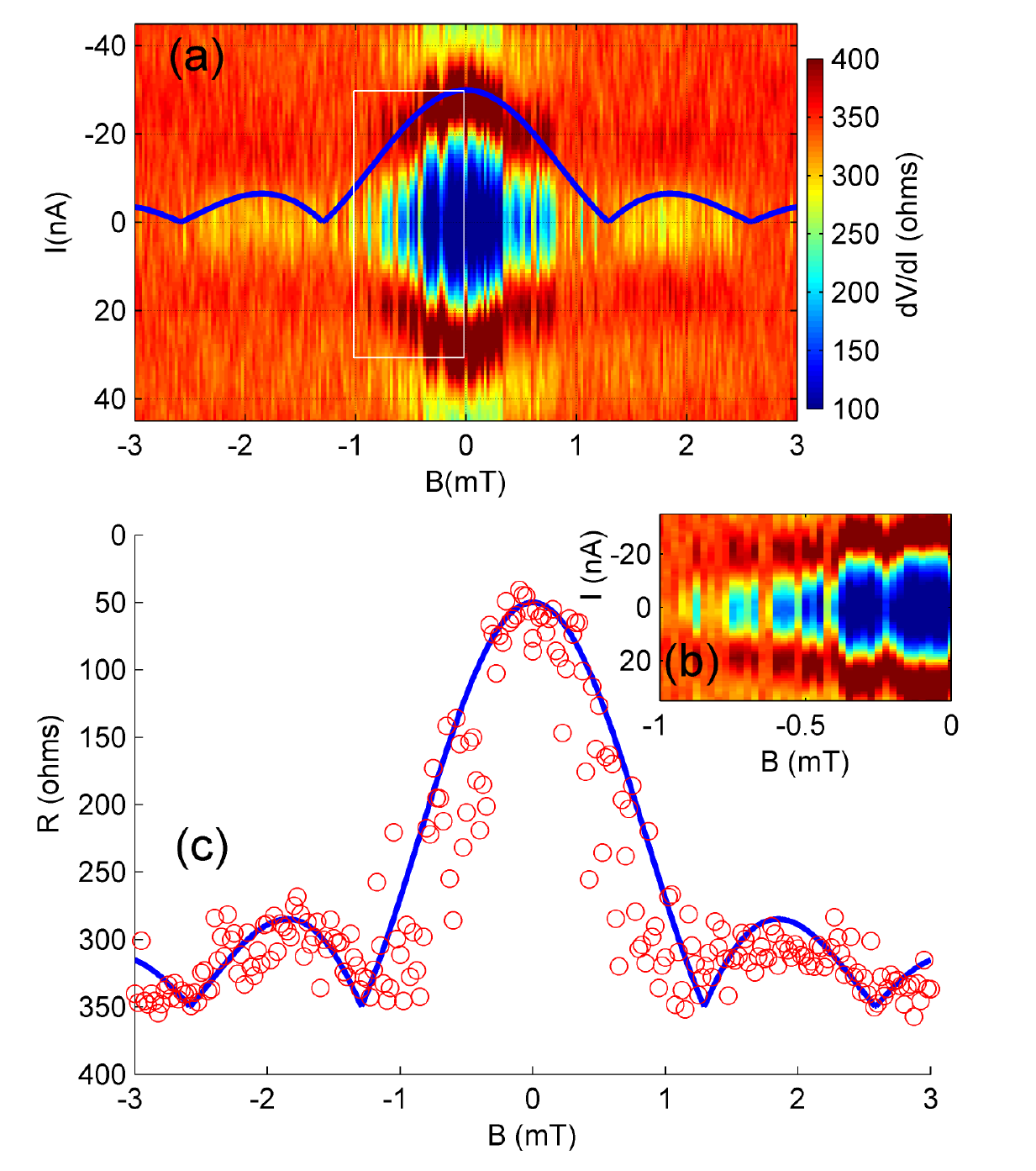}
\caption{(color online) (a) Colormap of the differential resistance $dV/dI(B,I)$ for the junction of width $L\approx 300$ nm presented in Fig.~\ref{fig:G0}. 
The data have been collected by decreasing the absolute value of the applied magnetic field. 
The dark blue area corresponds to the superconductive region. 
The superposed blue curve is a Fraunhofer interference pattern  given as a reference, corresponding to a total area $S_\mathrm{eff}= 1.6~\mu$m$^2$. (b) Enlargement of a region (white rectangle) of panel (a), evidencing additional fast oscillating patterns. 
(c) Residual resistance at $I$= 0 nA. The data are shown 	as open red circles. The  blue solid line appearing in panel (a) is also shown here, as a guide for the eye.}
\label{fig:FP}
\end{center}
\end{figure}

\begin{figure}[h]
\begin{center}
\includegraphics[width=0.9\linewidth]{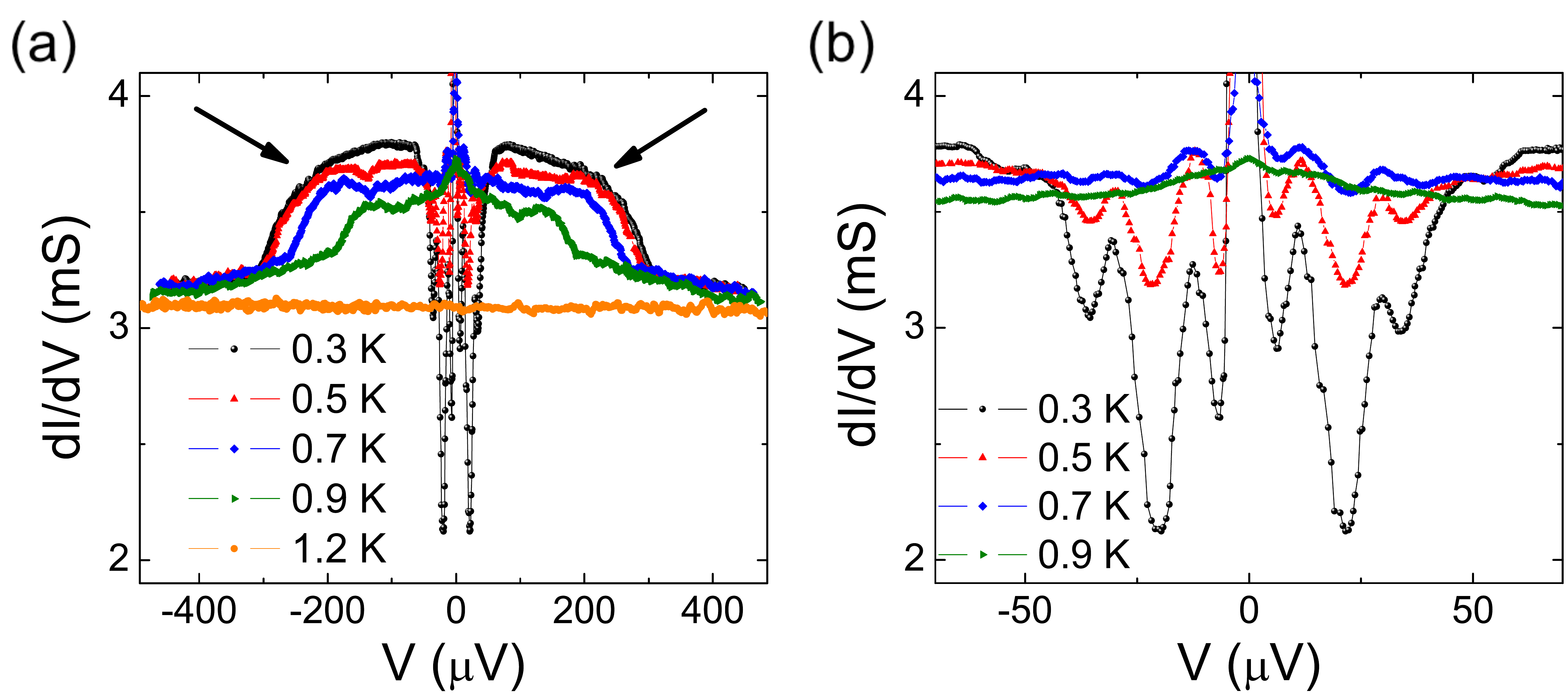}
\caption{(color online) (a) Measurements of conductance spectra $dI/dV (V)$ as a function of temperature on the $L \approx$ 200 nm junction. On a large scale the temperature modulation of the Al gap clearly appears, as indicated by the black arrows. Panel (b) is a zoom around zero voltage and shows that subgap peaks do not shift with temperature.}
\label{fig:spectra}
\end{center}
\end{figure}

The temperature dependence of the resistance of four junctions is presented in Fig.~\ref{fig:G0}(a). There is a kink at about 1~K monitoring the transition to superconductivity of the  Ti/Al contacts. The junction with $L \approx 600$ nm (green curve), appears to be going insulating. Not even in the smallest junctions with $L \approx 200$ nm (black curve) and $L \approx 300$ nm (red curve) is the transition to the superconducting state complete above 280 mK.  The precursory effects of the dissipationless conductance in these junctions will be discussed elsewhere. Here we show that, no matter that some residual resistance is present above 280 mK, these junctions attain phase coherence and Josephson conduction. Fig.~\ref{fig:G0}(b) shows the corresponding  $V(I)$ characteristics taken at $T = 280$ mK for the four junctions presented in panel (a). The $V(I)$ characteristic is practically ohmic  for the junction with $L \approx 600$ nm.  A Josephson critical current $I_c \approx 40$ and $10$ nA can be read off for the junctions with $L\approx 200$ nm and $300$ nm  respectively. The small steps at finite voltage below the Ti/Al gap can be attributed to localized subgap states (see below). The mean free path of graphene on hydrogenated SiC is estimated to be $\ell \sim 50$ nm from room temperature measurements and  the carrier density and mobility are only weakly modified from room temperature to liquid Helium temperature\cite{Speck2011APL,Jabakhanji2014PRBa}.
Thus even the smallest of the junctions should be diffusive.
Estimating the aluminium gap $\Delta \approx 100$ $\mu$eV from conductance $ dI/dV \:(V)$ measurements (see Fig. \ref{fig:spectra}(a)), we obtain a superconducting coherence length  $\xi_0 = \sqrt{\hbar D / \Delta} \approx 300$--$400$ nm, 
where $D$ is the diffusion constant. 
%
Hence our  junctions with $L  \le 300$ nm are in the crossover between the short and the long junction limit.  
A reduction of the Josephson critical current is expected in the long junction limit, as confirmed by the experiment.
%
%
%

For the $L\approx 200$ nm junction, the product $e I_c R_N$ $/\Delta$ $\approx 0.09$ is smaller than the theoretical estimate\cite{Dubos2001PRB} $e  I_c R_N/$ $\Delta \approx  0.66$ derived for  junctions  in which the Thouless energy is $\hbar D / L^2 \sim \Delta $. 
%
This is the upper limit of the difference, since it is calculated for  $I_c$  at $T  \approx 0.3$ $T_c^{Al}$ ($T_c^{Al}$ is the critical temperature of the electrodes).
The measured critical current density $J_c= I_c/W \approx$ $10$ nA $\mu$m$^{-1}$ at $L\approx 200$ nm  falls in the range of the values  reported in the literature, i.e. between 1 and 100 nA $\mu$m$^{-1}$, when the temperature is normalized to  $T_c$ of the electrodes\cite{CaladoNNANO2015,PopinciusPRB2012,Jeon2011PRB,Heersche2007Nature}.

One of the most direct evidence of a genuine Josephson coupling  is provided by a  modulation of $I_c$ with a  magnetic field $B$ which is applied perpendicularly to the substrate. The magnetic field dependence of the measured differential resistance for the $L \approx 300$ nm junction is reported in Fig.~\ref{fig:FP}(a). 
Experimentally, there is also a one to one correspondence between the magnitude of the Josephson critical current  and the small residual resistance. Consequently, a similar magnetic pattern is retrieved from the magnetic field  dependence of $dV/dI$ at  $I =0$, see panel (c). 
In panels (a,c), we have added for reference a conventional Fraunhofer interference curve:
$$
I_c \propto |\sin(\pi B S_\mathrm{eff}/\Phi_0)/ (\pi B S_\mathrm{eff}/\Phi_0)|$$
where $\Phi_0= hc/2e$ is the flux quantum. The one displayed here corresponds to an effective  area $S_\mathrm{eff}= 1.6$ $\mu$m$^2$. 
As the width of the junction is 4 $\mu$m and the length $L \approx $ 300 nm  one could infer
that the critical current density is homogeneous across the junction and that the physical area of the junction is involved, with the addition of some small penetration of the magnetic field in the Al/Ti contacts. However the sample, including the Al/Ti contacts, is a planar structure in the thin film limit and the  field penetration in the Al/Ti contacts is expected to be large. In fact, a conventional Fraunhofer pattern only  gives a qualitative envelope of a strongly oscillating pattern that is observed.

Fig.~\ref{fig:FP}, panel (b) is an enlargement of these low fields oscillations. They are quasi periodic, with a pseudo-period of $1$--$3$ Gauss which corresponds to a much larger effective area of  $\approx 3^2 \mu$m$^2$. 
To account for a more extended flux penetration in the contacts, we have numerically calculated the macroscopic magnetic vector potential profile by solving the London equation, following a method originally employed by Rosenthal {\it et al.}\cite{rosenthal,tafuri,arpaia} for planar junctions.
The $I_c$  oscillations with the  period of 3 Gauss and their enveloppe can be  reproduced in this way by choosing a Pearl penetration length within the contacts $\lambda=1$ $\mu$m and a non uniform critical current density across the junction.
However, a macroscopic London picture is still not fully convincing as  the aperiodicity of the fast oscillating pattern, the residual resistance (see Fig. 3(c))  and some (fully reproducible) dependence on the sweeping of the applied magnetic field imply additional phenomena which cannot be accounted for within the macroscopic Rosenthal model. Most likely, vortex pinning in the Al/Ti contacts takes also place, as many impurities and pinning centers are expected to diffuse in the contacts during  the deposition process. The fast oscillations in the pattern could be due to a  non-uniform critical current density distribution\cite{Barone,hart2014Nature}. A more detailed account on the magnetic field dependence of $I_c$ will be given elsewhere\cite{noi}.

When investigating the behavior of the Al/Ti  contacts further, by monitoring the conductance measurements $dI/dV (V)$ for different temperatures reported in Fig. \ref{fig:spectra}, we can identify the Al/Ti gap, with a marked temperature dependence (panel (a)), and few subgap structures which give rise to the steps in the $V(I)$ characteristics at finite voltage, see Fig. \ref{fig:G0}(b). Temperature seems not to have any effect on these subgap structures, as shown in Fig. \ref{fig:spectra}(b). This rules out the possibility that they originate from multiple Andreev reflection (MAR)\cite{DuPRB2008}. This can be explained by considering that the interfaces with the contacts are expected to be rather rough. Poor transmission between the metallic aluminium contacts and the underlying graphene layer, responsible for a sizeable contact resistance, has also been found in the case of diffusive graphene transis\-tors\cite{Wu2012NL}. Possible origin of the subgap resonances  could be resonant tunneling via localized states or even edge states in the graphene barrier. This feature will be investigated further in the future.


Uncovering the structure of the Al/Ti contacts in these devices is of the utmost relevance. Graphene and the Ti/Al contacts have a large work function difference and graphene may become $n$-doped by charge transfer from the Al/Ti contact\cite{Giovanneti2008PRL}.
As graphene on the hydrogenated SiC interface is intrinsically $p$-doped, devices could be engineered in the form of $n$-$p$-$n$ junctions. A Fabry-P\'erot resonating transmission which has been found in perfect ballistic $n$-$p$-$n$ junctions\cite{Allen2015} can be ruled out here, as our junctions are diffusive and the interfaces with the contact have low transparency. Still, this may be improved, as the mobility of G/SiC is known to increase strongly when the carrier concentration decreases.

%
%
%
%

\section{Conclusion}

Looking from the point of view of the graphene community, efforts have been mostly driven towards the realization of samples achieving a) a ballistic transport regime to exploit the unique properties of Andreev reflection in graphene\cite{shalom2015NatPhys,MizunoNCOM2013,CaladoNNANO2015,Baringhaus2014Nature}; 
b)  a more efficient control of charge density, with the possibility of reaching  the neutrality point through electrostatic gate\cite{DeonPRL2014,ChoiNCOM2013}.
%
%
The possible use of graphene junctions in real circuits is hindered, however, by the substantial limit of a technology that cannot produce a large number of junctions patterned on a single graphene flake with similar properties, thus controlling the electrodynamics  and functionality of each of them, as well as of the global circuit. 

This work demonstrates  the Josephson conduction at subKelvin temperature, in various co-planar  structures on one single monolayer of graphene    and  patterned Al superconducting contacts. Graphene is grown by CVD on SiC. The Josephson coherence appears to be highly reproducible,  isotropic with respect to the substrate orientation and easily scalable with present standard lithographic techniques. These results have an immediate applicative impact opening up  to a systematic  comparative study  where single constructive parameters of the  graphene junctions  can be selectively changed and barriers can be appropriately engineered.

%
%
%
%
\section{Acknowledgements}

Discussions with Piet Brouwer and Victor Rouco Gomez are gratefully acknowledged. Work supported by PICS CNRS-CNR 2014-2016 "Transport phenomena and Pro\-xi\-mi\-ty-induced Super\-conduc\-ti\-vi\-ty in Graphene junctions", Swedish Foundation for Strategic Research (SSF) under the project "Graphene based high frequency electronics", FIRB  "HybridNanoDev RBFR1236VV"  (Italy) and by EU FP7, under grant agreement no 604391 Graphene Flagship.

\bibliography{library_apl} 
\end{document}